\documentclass[shortnote]{jpsj3}
\usepackage{txfonts}
\usepackage{here}
\newcommand{\tc}{$T_{\mathrm{c}}$}
\newcommand{\lrp}{LaRu$_4$P$_{12}$}
\newcommand{\hc}{$H_{\mathrm c2}$}
\newcommand{\tone}{1/$T_1$}
\newcommand{\tonet}{1/$T_1T$}
\renewcommand{\figurename}{Fig.}
\newcommand\figref[1]{Fig.~\ref{fig:#1}}
\newcommand\figureref[1]{Figure~\ref{fig:#1}}

\title{Magnetic Field Effect on $s$-wave Superconductor LaRu$_4$P$_{12}$ Studied by $^{31}$P-NMR}

\author{Katsuki Kinjo$^{1}$\thanks{kinjo.kats
uki.63v@st.kyoto-u.ac.jp}, Shunsaku Kitagawa$^1$, Yusuke Nakai$^{1}$\thanks{Present Adress: Department of Physics, Hyogo University, Hyogo 678-1297, Japan}, Kenji Ishida$^1$, Hitoshi Sugawara$^{2}$\thanks{Present Adress: Department of Physics, Kobe University, Hyogo 657-8501, Japan}, Hideyuki Sato$^2$}
\inst{$^1$ Department of Physics, Kyoto University, Kyoto 606-8502, Japan \newline
$^2$ Department of Physics, Tokyo Metropolitan University, Tokyo 192-0397, Japan 
} 

\abst{We have performed $^{31}$P-NMR measurements on the $s$-wave superconductor \lrp\ to investigate the magnetic field effect of the nuclear spin-lattice relaxation rate \tone\ on a conventional full-gap superconductor. With increasing magnetic field, the Hebel-Slichter peak immediately below \tc\,in $1/T_1$ was suppressed, and the magnetic field dependence of \tone\ at 0.8~K, well below \tc, was proportional to $H^2$.
These behaviors can be fully understood by the orbital pair-breaking effect in a single-band $s$-wave superconductor.}

\begin{document}
\maketitle

Superconducting (SC) states near an upper critical field \hc\ have attracted much attention because an exotic SC state, such as the Fulde-Ferrell Larkin-Ovchinnikov (FFLO) state\cite{Fulde,Larkin}, is expected when the Pauli pair-breaking effect predominates over the orbital pair-breaking effect. 
In fact, there are some reports that the FFLO phase or Q phase (the coexistence phase of the FFLO state and spin density wave state) seems to be realized near \hc\ in heavy fermion superconductors and organic superconductors\cite{CeCoIn5,BEDT,CeCu2Si2,UPd2Al3}. 
Recently, it has been recognized that $1/T_1$ measurement around $H_{\rm c2}$ is a valuable method for studying the FFLO state because the enhancement of $1/T_1$ has been observed in $\kappa$-(BEDT-TTF)$_2$Cu(NCS)$_2$ and CeCu$_{2}$Si$_{2}$, and the enhancement of $1/T_1$ suggests the formation of the FFLO state.
In contrast to extensive NMR studies of unconventional superconductors, there are few NMR measurements near \hc\ on conventional $s$-wave superconductors, and thus a detailed NMR study near \hc\ for conventional superconductors is important to understand the magnetic field effect on superconductivity. 

In the case of a conventional superconductor, we adopted filled skutterudite \lrp\ with a moderate \tc\ and \hc.
\lrp\ has a cubic symmetry $Im\bar{3}$ space group. 
The SC transition temperature \tc\ of LaRu$_4$P$_{12}$ is 7.2~K, and the upper critical field \hc\ is 3.4~T.\@ 
From various measurements\cite{SC,SubmeV}, \lrp\ is considered to have an $s$-wave SC symmetry. 
In addition, the temperature dependence of \hc\ is consistent with the Werthamer-Helfand-Hohenberg theory\cite{WHH}, indicating that the orbital pair-breaking effect is dominant. 
Therefore, \lrp\ is a good candidate for investigating the magnetic field effect on a conventional superconductor with orbital pair-breaking. 

In this short note, we perform $^{31}$P-NMR measurements on LaRu$_4$P$_{12}$. The Hebel-Slichter (HS) peak in the nuclear spin-lattice relaxation rate $1/T_1$, which was observed at 1~T, was suppressed with increasing magnetic field.
This suppression originates from the Volovik effect for conventional $s$-wave superconductors.
In addition, the magnetic field dependence of \tone\ at 0.8~K was proportional to $H^2$, which is quite consistent with the expected behavior in superconductors in which the single-band orbital pair-breaking effect is dominant.

\begin{figure}[t]
\centering
\includegraphics[width=\linewidth]{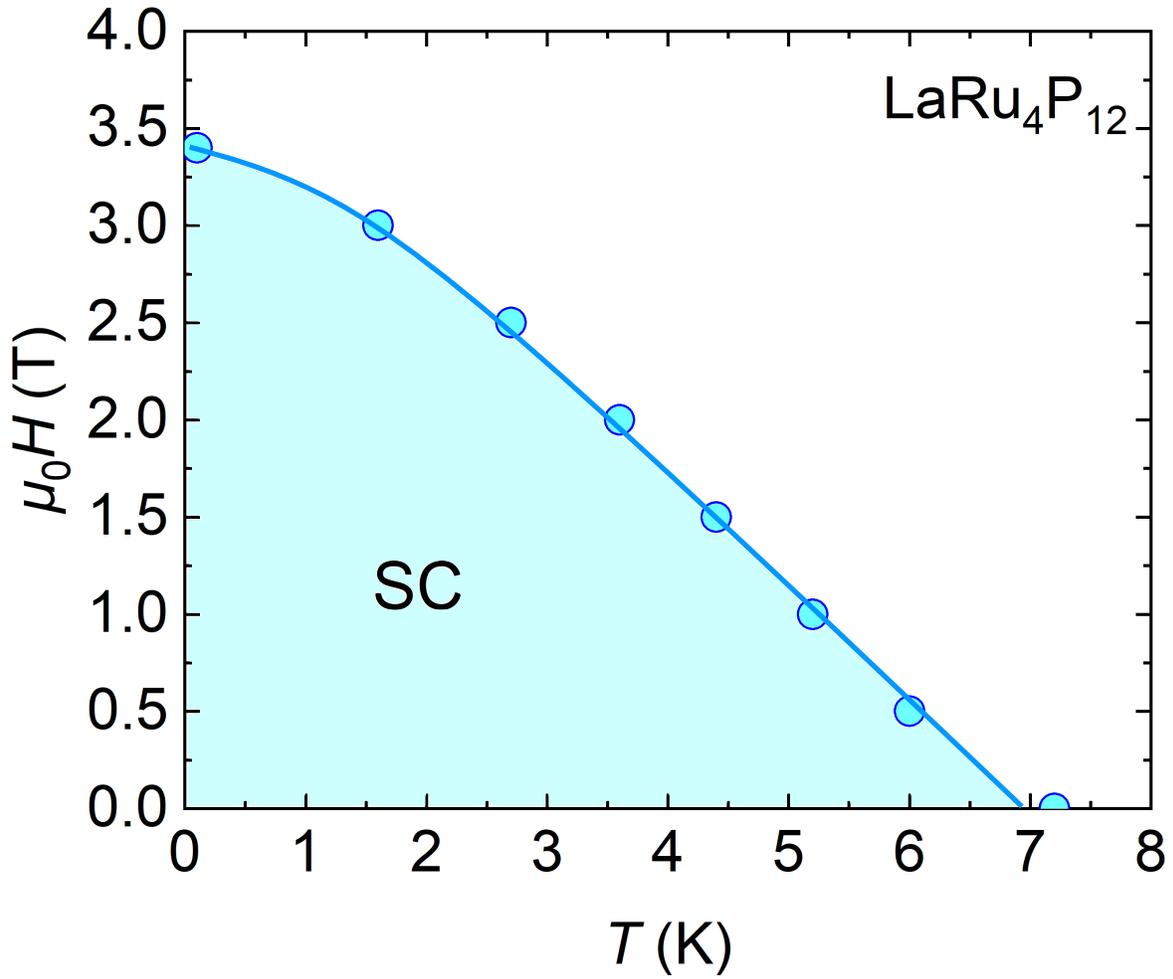}
\caption{(Color online)$H$-$T$ phase diagram of \lrp. \tc s are determined by $\chi_{\mathrm{AC}}$ measurements with in-situ NMR coils.}
\label{fig:phasediagram}
\end{figure}

\begin{figure}[t]
\includegraphics[width=\linewidth]{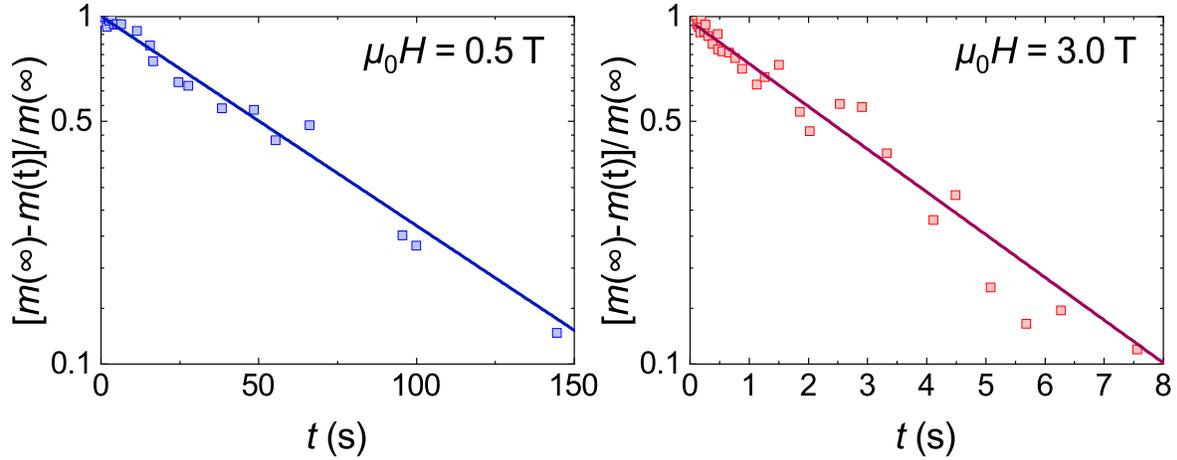}
\caption{(Color online)Relaxation of the magnetization of $^{31}$P nuclei (left) at 0.8~K and 0.5~T and (right) at 0.8~K and 3~T. Solid lines are the following equation: $[m(\infty)-m(t)]/m(\infty) =A \exp{(-t/T_1)}$.}
\label{fig:recovery}
\end{figure}

\begin{figure}[t]
\includegraphics[width=\linewidth]{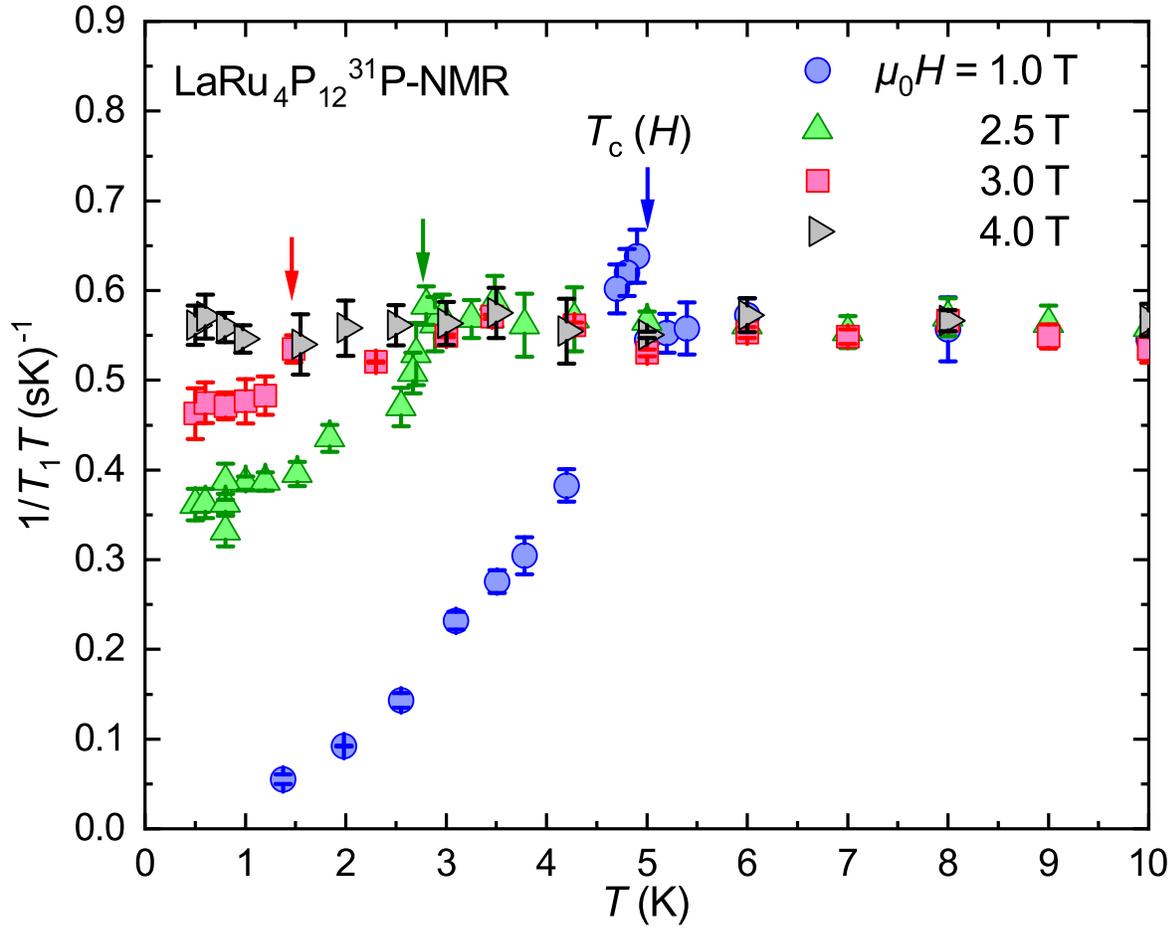}
\caption{(Color online)Temperature dependence of \tonet\ at several magnetic fields. The arrows indicate \tc(\it{H}).}
\label{fig:t1loglog}
\end{figure}

A single crystal of \lrp\ was synthesized by the Sn-flux method \cite{SC} and was powdered for NMR measurements to obtain a large surface area.
The field dependence of $T_{\rm c}(H)$ [$\mu_0H_{\rm c2}(T)$] was obtained by AC-susceptibility measurements using NMR coils.
As shown in \figref{phasediagram}, the observed $T_{\rm c}(0)$ (= 7.2~K) and $\mu_0H_{\rm c2}(0)$ (= 3.4~T) are consistent with a previous report\cite{SC}.
A conventional spin-echo technique was used for NMR measurements in a temperature range from 0.2 to 10~K and magnetic field range from 0.5 to 4.0~T.
Low-temperature measurements below 1.5~K were carried out with a $^3$He-$^4$He dilution refrigerator, in which the sample was immersed into the $^3$He-$^4$He mixture to prevent radio-frequency heating during measurements.  
1/$T_1$ of $^{31}$P nuclei (nuclear spin $I$ = 1/2 and nuclear gyromagnetic ratio $^{31}\gamma/2\pi = 17.235$~MHz/T) was measured using the saturation recovery method and was uniquely determined with a single component in the entire measurement region, even near \hc\, as shown in \figref{recovery}.

\figureref{t1loglog} shows the temperature dependence of \tonet\ at 1, 2.5, 3, and 4~T. In the normal state, \tone\ is proportional to temperature, which is well known as the Korringa behavior ($T_1T=\mathrm{const.}$). 
This indicates that LaRu$_4$P$_{12}$ is a normal metal without strong magnetic fluctuations below 10~K.
Below \tc, \tone\ deviates from the Korringa behavior. 
At 1~T, \tonet\ shows a clear peak, the so-called HS peak\cite{HSpeak}, immediately below \tc({\it H}), and decreases exponentially at low temperatures, which is quite consistent with the expected behavior in full-gap $s$-wave superconductors. Note that the temperature dependence of \tone\ deviates from the exponential curve below 3~K.
The detailed NMR study at low magnetic fields has been reported in Ref. [11].
With increasing magnetic field, the HS peak was suppressed and there was no clear peak at 2.5 and 3~T.
In addition, at 3~T, the decrease in $1/T_1$ below $T_{\rm c}$ was suppressed and a tiny deviation from the Korringa behavior was observed.
The suppressions of the HS peak originate from the Doppler shift effect of the quasiparticle excitation, known as the Volovik effect in $s$-wave superconductors\cite{volovik}; this effect was first pointed out in CaPd$_2$As$_2$ \cite{CaPd2As2}. 
In contrast to superconductors which are expected to have an FFLO phase near $H_{\rm c2}$, there was no enhancement of $1/T_1$ near $H_{\rm c2}$(0).

\begin{figure}[H]
\includegraphics[width=\linewidth]{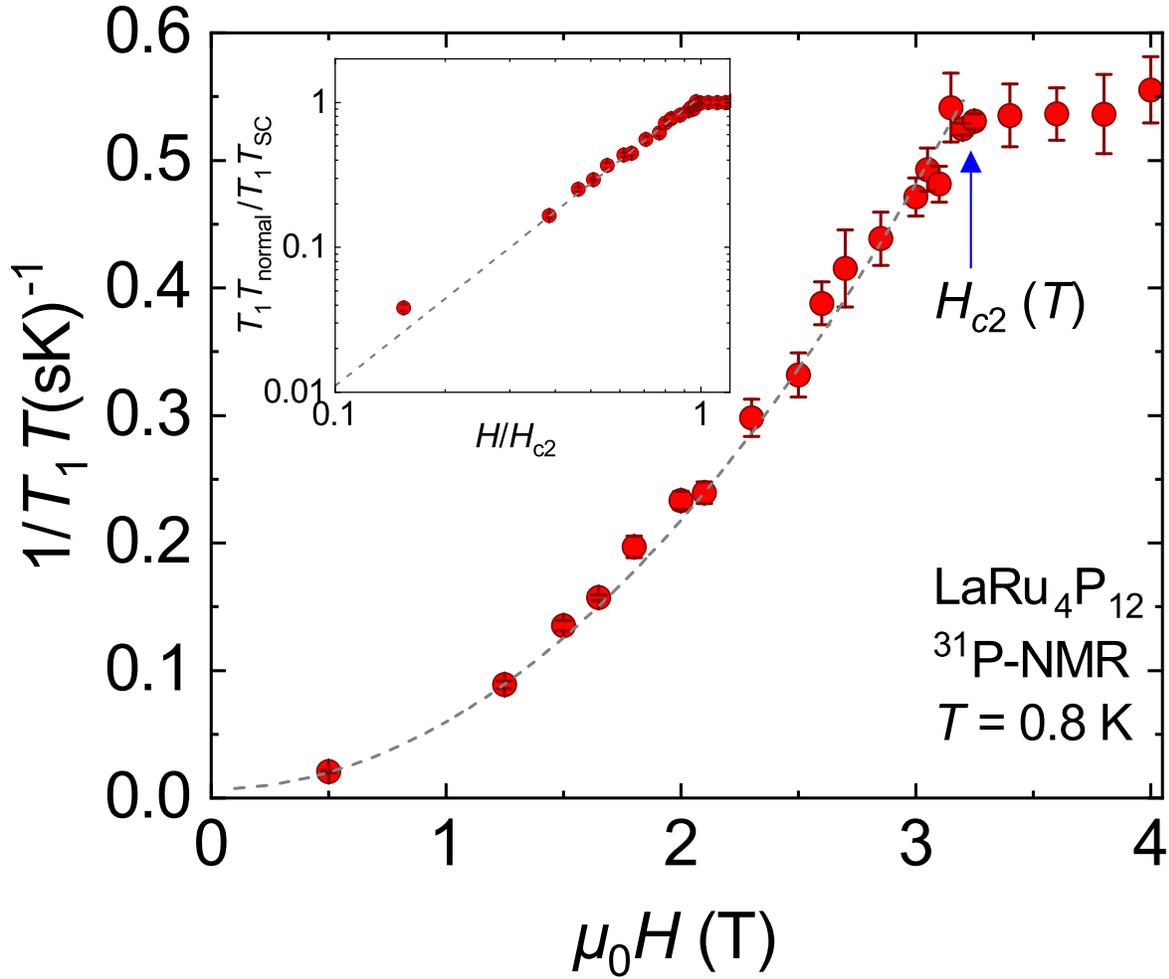}
\caption{(Color online)The {\it H}-dependence of \tonet\ at the P site at 0.8~K. The break line is $a+bH^2$. (inset) The double-logarithmic chart of normalized {\it H}-dependence of \tonet. The break line is $(H/H_{\mathrm c2})^2$.}
\label{fig:t1thdep}
\end{figure}

We investigated how \tonet\ changes with respect to the magnetic field at low temperatures.
\figureref{t1thdep} shows the magnetic field dependence of \tonet\ at 0.8~K, where \tonet\ is determined with the vortex state.
\tonet\ increases with increasing magnetic field and smoothly connects with the normal-state value. 
Again, there was no enhancement of \tonet\ below $H_{\rm c2}$, which is different from FFLO superconductors.
Above \hc, \tonet\ remains constant with the magnetic field.
As shown in the inset of Fig.~\ref{fig:t1thdep}, \tonet\ in the SC state was proportional to $H^2$.
This is in contrast to a line-node superconductor, in which quasiparticles are induced at nodes and the density of states (DOS) is proportional to $\sqrt{H/H_{\mathrm c2}}$, resulting in a $H$ linear dependence of \tonet\cite{tei}.
The experimental results can be fitted with $a+bH^2$.
Here, $a$ is the value of $1/T_1$ at 0~T, and $b$ is the coefficient.
As shown in the inset of \figref{t1thdep}, \tonet\ below 1~T did not follow $H^2$ behavior, most likely because of some impurities in the sample.
In Type II superconductors, the quasiparticle DOS in a magnetic field is proportional to the number of vortex cores and \tonet\ is proportional to the square of the DOS.
In full-gap superconductors, the induced quasiparticle DOS is proportional to $H$; thus, $1/T_1T$ is proportional to $H^2$.
All experimental results are consistent with a full-gap $s$-wave superconductor with orbital pair-breaking effect. 
However, theoretical calculation suggests that the quasiparticle DOS of $s$-wave full-gap superconductors under a low field is expressed by the following equation: $N_{\mathrm{loc}}(H,E=0)/N_0 \sim H/0.8H_{\mathrm{c2}}$\cite{DOS}, where $N_{\mathrm{loc}}$ and $N_0$ are localized quasiparticle DOS and DOS in the normal state, respectively, and the coefficient of $(H/H_{\rm c2})^2$ is smaller than 1 near \hc. 
In the present study, $N_{\rm loc}/N_0 \sim H/H_{c2}$ in the field region of $H/H_{\rm c2} > 0.4$, and the $H$ dependence suggested theoretically was not observed.

In conclusion, we performed $^{31}$P-NMR measurements over a wide magnetic field range on the conventional $s$-wave superconductor \lrp.
The suppression of the HS peak under a magnetic field and $H^2$ dependence of $1/T_1$ are consistent with a full-gap $s$-wave superconductor with orbital pair-breaking effect. In addition, $1/T_1$ does not show any enhancement below $T_{\rm c}$ near $H_{\rm c2}(0)$. This behavior is in contrast with that of FFLO superconductors.
This detailed NMR study of a conventional superconductor is useful for understanding the effect of a magnetic field on superconductivity.

\begin{acknowledgment}
This work was partially supported by the Kyoto Univ. LTM Center and Grant-in-Aids for Scientific Research (KAKENHI) (Grant Numbers JP15H05882, JP15H05884,  JP15K21732, JP15H05745, JP17K14339, and JP19K14657). The authors would like to thank T. Okuno, G. Nakamine, M. Manago, R. Kotai, A. Ikeda, Y. Maeno, and S. Yonezawa for valuable discussions. We would like to thank Editage (www.editage.jp) for English language editing.
\end{acknowledgment}

\end{document}